Preliminary Examination of Guardian Cap Head Impact Data Using Instrumented Mouthguards

Kristen G Quigley[1], Dustin Hopfe, MS, LAT, ATC[1], Madison R Taylor[1], Philip Pavilionis, MS, ATC[1], Vincentia Owusu-Amankonah[1], Arthur Islas, MD[2], Nicholas G Murray, PhD[1]*

[1]Neuromechanics Laboratory, School of Public Health, University of Nevada, Reno, Nevada 89557, USA

[2] School of Medicine, University Of Nevada, Reno, Nevada, 89557, USA

*Corresponding author at:

Nicholas G. Murray

Neuromechanics Laboratory

Department of Kinesiology

School of Public Health,

University of Nevada, Reno

1664 N Virginia Street m/s 0274, Reno, NV 89557, USA.

Tel: (775) 682-8347;

Fax: (775) 784-1340.

E-mail address: nicholasmurray@unr.edu

**Abstract**

*Purpose*

The objective of this study is to present preliminary on-field head kinematics data for NCAA Division I American football players through closely matched pre-season workouts both with and without Guardian Caps (GCs).

*Methods*

42 NCAA Division I American football players wore instrumented mouthguards (iMMs) for 6 closely matched workouts, 3 in traditional helmets (PRE) and 3 with GCs (POST) affixed to the exterior of their helmets. This includes 7 players who had consistent data through all workouts.

*Results*

There was no significant difference between the collapsed mean values for the entire sample between PRE and POST for peak linear acceleration (PLA) (PRE=16.3±2.0, POST=17.2±3.3Gs; p=0.20), Peak Angular Acceleration (PAA) (PRE=992.1±209.2, POST=1029.4±261.1rad/s2; p=0.51 and the total amount of impacts (PRE=9.3±4.7, POST=9.7±5.7; p=0.72). Similarly, no difference was observed between PRE and POST for PLA (PRE=16.1±1.2, POST=17.2±2.79Gs; p=0.32), PAA (PRE=951.2±95.4, POST=1038.0±166.8rad/s2; p=0.29 and total impacts (PRE=9.6±4.2, POST=9.7±5.04s; p=0.32) between sessions for the7 repeated players.

*Conclusion*

These data suggest no difference in head kinematics data (PLA, PAA and total impacts) when GCs are worn. This study suggests GCs are not effective in reducing the magnitude of head impacts experienced by NCAA Division I American football players.

**Introduction**

Brain injuries have been heavily studied in recent years, with a particular emphasis on mild traumatic brain injuries (mTBI), or concussion. Sports-related concussion (SRC) in particular remains a vital public health issue that affects participants of all ages and levels of sport. For collegiate athletes specifically, SRC represents approximately 6% of all athletic injuries, with American football serving as the largest contributor to this statistic (1). On top of these already troubling head injury statistics, many athletes experience a phenomenon known as repetitive head impact (RHI), which is defined as multiple powerful blows to the head that are not significant enough to result in the clinical diagnosis of a concussion (2). RHI is particularly common among American football players due to the repetitive blows incurred on each subsequent play for the majority of the players. Previous research suggests that frequent exposure to RHI may lead to changes in white matter connectivity and decreased activation of the dorsolateral prefrontal cortex, which is the brain area primarily responsible for executive function and decision making (2-4). The combination of these factors presents a clear need for interventions to better protect athletes of all domains, with a particular emphasis on American football players.

Research on American football has seen many positive innovations over the years, particularly with the implementation of instrumented mouthguards (iMM) to collect data regarding changes in rotational and linear kinematics. Instrumented mouthguards can be difficult to use due to the need for athletes to be gentle with the hardware (i.e., avoiding chewing), however, they are a considerable advancement and provide comparable data to the traditional helmet-based systems (5). When used properly with a custom dental scan for fitting and routine maintenance, iMMs can measure the amount of impacts each user experiences during the recording session. That said, best scientific practices suggest advanced filtering techniques and substantial video verification. This does limit the immediate on-field applications, however iMMs provide a comfortable and affordable way to accurately track head impacts to aid researchers in answering critical questions about preventative approaches for SRC and/or RHI.

The most straightforward course of action to better protect American football players from the effects of RHI or SRC would be to improve helmet technology. Though RHI research is relatively new, researchers have been attempting to create helmets better at attenuating the forces applied to the head for years. The push for better helmet design can involve altering the shell, inner padding or adding additional padding to the exterior of the existing helmet (6). One of the first innovations of this kind was the ProCap (Protective Sports Equipment Inc, MD), which was released in 1989 and consisted of a hard-shell cover that affixed to the exterior of a traditional football helmet (6). ProCap was endorsed by some American football players in the National Football League (NFL), however the ProCap's popularity dwindled when the primary NFL helmet manufacturer of that time, Riddell, revoked the warranty from their helmets if they had been modified with a ProCap (6).

In 2011 the National Operating Committee on Standard for Athletic Equipment (NOCSAE) backed Riddell's decision from years prior, stating that the addition of components to athletic equipment or modification of the original equipment voids the warranty and safety certifications of the product (6). This regulation was later revisited by NOCSAE and overturned, so long as the company wishing to design "helmet add-ons" tested the product themselves and became responsible for the warranty of the equipment (7). This change in legal proceedings was what allowed for the current market leader, Guardian Sports, to create the Guardian Cap (Guardian Sports, Peachtree Corners, GA) in 2015.

The Guardian Cap (GC) is currently the most popular external helmet add-on which aims to attenuate the head impact magnitude experienced by American football players. The GC follows a similar design to the previously mentioned ProCap, however it is classified as a soft-shell cover (8). The GC popularity is in part due to heavy endorsement from the NFL, who employed GC during their 2022 pre-season training. The NFL stated that they saw a 50% reduction in SRC when compared to the previous averages from 2018, 2019 and 2021, however the data behind this claim has never been published (9). Other studies have tested the effectiveness of the GC, however the majority of these studies only used laboratory helmet impact testing, such as vertical drop testing (10). Only one study exists with on-field measurements to date, which utilized instrumented mouthguards to collect angular and rotational head kinematics from 5 National Collegiate Athletic Association (NCAA) Division I American football players (11). This study utilized 13 workouts worth of on-field data collected in 2019 from 5 linebackers wearing traditional helmets and compared it to 14 workouts worth of on-field data collected in 2021 from a unique set of 5 linebackers wearing GC (11). While prior research is sparse, it suggests that the GC does not alter or reduce the head kinematics data, but additional data is needed from a larger and more robust cohort to confirm these findings.

To address the endorsement from the NFL and need for larger on-field data validations, this paper aims to establish preliminary angular and rotational head kinematics data for NCAA Division I American football players through closely matched practice sessions both with and without GC. Consistent with limited previous literature, the researchers hypothesize that GCs will have no significant reduction on the peak linear acceleration (PLA), peak angular acceleration (PAA) experienced by American football players during preseason workouts.

**Methods**

*Participants*

42 NCAA Division I male American football players (average age=20±1 year) participated in this study (Table 1). This included 7 players who participated in the six practice sessions and had complete data across (Table 2). This resulted in 83 participants, with 7 repeated players. Participants were recruited from the same American football team if they wore dental scanned and fitted instrumented mouthguards (iMM) from Prevent Biometrics to record head impacts (3200 Hz; Prevent Biometrics, Edina, MN) during pre-season workouts. Each iMM contains a triaxial accelerometer and gyroscope to measure linear and rotational kinematics. The football players' position varied (OL = offensive lineman, DL = defensive lineman, RB = running back, TE = tight end, LB = linebacker), but all were identified by the coaching staff as having the potential to be high dose players (Table 1). Fourteen percent of the players were deemed starters, 40% were rotational players and 46% were scout team players as denoted by the coaching staff (Table 1). The players were allowed to choose the brand of their traditional helmet; 83% of players wore the Riddell SpeedFlex (Riddell Sports Group, Rosemont, Illinois, USA), 3% of players wore the Vicis Zero2 (Vicis, Seattle, WA, USA), 14% of the players wore the Schutt F7 2.0 (Schutt Sports, Litchfield, IL, USA). The players used the same helmet for each of the 6 recorded workouts.

Participants were excluded if they did not adhere to the iMM compliance standards (i.e., excessive chewing) and if they had an injury that precluded them from practicing the day of the selected practices. Of the 85 total participants, 2 participants had been diagnosed with a concussion within 6 months of the start date of the study but were medically cleared to return-to-sport. 1 participant was diagnosed with a concussion during the study and was then excluded.

All participants agreed verbally and with written informed consent to participate in the study. This study was approved by the Institutional Review Board of the respective university (IRB number: 1757959-5) and in accordance with the Declaration of Helsinki.

*Practice Plans and Guardian Cap*

The research team selected 6 (3 pre-GC [PRE] and 3 post-GC [POST]) nearly identical practice plans that contained a similar level of hitting and time for drills. The practices lasted approximately 2 hours with: 15 minutes of tackling drills, 25 minutes of individual position drills (which included a tackling circuit), 60 minutes of team drills using THUD (initiation of contact at full speed with no predetermined winner, but no take-down to the ground) tackling, and 20 minutes of team tempo play TAG (tackling to the ground). Practices varied slightly but no practices were denoted as scrimmages.

The PRE practices occurred within 24 hours of one another, while the POST practices occurred 5 days between each session. The last PRE practice was two days before the first POST practice session. The days between practices varied during the POST due to the sports season moving from the acclimation period and preseason to the regular season (12). During the regular season, full contact TAG is limited to 1-full contact day per week per NCAA recommendations (12). All participants in the POST had GCs that were fitted by the equipment staff and verified to be working condition before each session.

*Instrumented Mouthguards*

Before the beginning of the season, all participants were dental scanned, and a custom mouthguard was created by Prevent Biometrics. Based on preliminary data using the industry's highest standards for head impact verification, the Prevent mouthguards are 95% accurate in the detection of true positive impacts (13-15). When a participant incurs a blow ≥5g PLA, the sensor collected data for 16 pre- and 144 post-trigger samples. Each participant was instructed to always wear the iMM and to refrain from placing it in areas that may cause breakage. All the head impacts were filtered using the custom algorithm from Prevent Biometrics, which is approximately 95% accurate in the detection of a head impact event (13-15). In addition, all reported true positive head impacts were video verified using three camera (4k/HD AG-UX180 Handheld Camcorder, Panasonic, Kadoma, Japan) angles (both end zones and the 50-yard line) on a full-size practice field. One person, with extensive NCAA Division I film review experience, video verified all the true positive head impacts. If any impacts were not video verified, that impact was removed from the overall analysis. PLA, PAA and the total amount of impacts were analyzed across the 6 practice sessions. Each impact was considered its own individual event and ensemble averaged.

*Statistical Analysis*

All data were examined for influential skewness and kurtosis. The head kinematics data were normally distributed, and an initial independent samples t-test determined no statistical difference between the PRE practices and POST practices in all head kinematic data (Table 3). Thus, we collapsed the PRE sessions and the POST sessions and performed one-way ANOVAs for PLA, PAA and total impacts. In addition, the 7 players who had complete datasets were independently analyzed using repeated measures ANOVA across all six sessions for PLA, PAA and total impacts. Follow up paired samples t-tests were completed in the event of a significant time effect for the entire sample (Table 3), as well as independent samples t-tests for the 7 consistent players (Table 4). The alpha was set at 0.05 *a priori*.

**Results**

*PRE and POST Values for PLA, PAA and Total Impacts*

The mean values for the average PLA, PAA and total impacts for the collapsed PRE workouts and for the collapsed POST workouts using data from both the entire sample and only the seven consistent players can be found in Table 5. The mean values for the average PLA, PAA and total impacts for each individual workout can be found in Table 6.

*One-way ANOVA PLA, PAA and Total Impacts Across the Entire Sample*

A one-way between-subjects ANOVA revealed no significant difference between the collapsed mean PRE and POST PLA ($F(1,83) = 1.67$, p=0.20), PAA ($F(1,83) = 0.44$, p=0.51) and total impacts ($F(1,83) = 0.13$, p=0.72) These data suggest that there was no overall significant difference between the head kinematics for all the players before and after GC implementation.

*One-way Repeated Measures ANOVA Across the Seven Consistent Players*

A one-way within-subject repeated measures ANOVA revealed no significant differences between the collapsed mean PLA ($F(1,6) = 1.16$, p=0.32), PAA ($F(1,6) = 1.36$, p=0.29), and total impacts ($F(1,6) = 0.03$, p=0.88) pre- and post-Guardian Cap implementation (Table 5). These data suggest that in players who participated in all six practice sessions, the GC did not significantly alter the head kinematics data.

*Discussion*

The purpose of this study was to evaluate the effect that GCs have on the head kinematics during similar on-field practices in American Football using iMMs, to address the growing concern around RHI and brain injury risk in sport. To the authors' best knowledge, this is the largest on-field measurement of iMMs using GCs with an analysis of 809 unique video verified head impacts across varying players, different positions, and various playing time roles. This study was able to obtain multiple practices worth of data from the same players throughout a single season, which is something other studies have not been able to accomplish. The major outcome of this study is that the GCs did not reduce or attenuate the PLA, PAA or alter the total amount of head impacts between the 6 closely matched practice sessions. In addition, no change in any variables was noted after implementation of the GCs when comparing the same 7 athletes across the 6 practice sessions. While it has been reported by media outlets, which are not peer-reviewed nor have the data been made publicly available, that GCs greatly reduce the incidence of concussion, our results suggest they do not reduce the overall head kinematic data on the field (9,16). While no direct link exists between head kinematic data and concussions, the total amount of head impact exposure (i.e. frequency and magnitude) are generally higher the days leading up to a concussion diagnosis (17,18). The results of the current study do not suggest that the GCs reduce the overall head impact exposure, however, concussions were not tracked during this study.

These results support prior laboratory research that GCs do not significantly reduce head kinematic data during use (10). The majority of the research published on GCs is laboratory research, which has consisted of dropping the GCs from various heights and measuring the force the helmet experiences when hitting the ground. Laboratory testing is highly recommended for preliminary investigations on safety concerns when implementing new technology to sport, however it lacks ecological validity. The most effective line of testing is to use the technology in a real-world setting, which is what the present study aimed to do by using GCs during closely matched practices. When directly comparing our data to previously published laboratory examinations of GCs, the PLA and PAA are similar to existing research (19).

To our knowledge, there is only a single study using on-field data to validate the use of GCs (12). The study done by Cecchi et al. (2022) utilized instrumented mouthguard data from 5

Division I linebackers collected over the course of two separate years of training, which provided data from 13 practices using traditional helmets in 2019 and 14 practices using GCs in 2021. The 5 players used in the earlier study were not consistent throughout the two recorded seasons. Though our study does present significant differences in methodology, as we analyzed data from a larger sample set over the course of a single season, the study by Cecchi et al. (2022) is currently the only published work available for comparison. The present study found no significant differences in PAA, PLA, and total impacts as did Cecchi et al. (2022). They also reported no significant differences in Diffuse Axonal Multi-Axis General Evaluation (DAMAGE) or Head Acceleration Response Metric (HARM), however these variables were not evaluated in the present study. The combination of results from Cecchi et al. (2022) and the present study strongly suggests that GCs are not effective in reducing the amount and the magnitude of head kinematics experienced by collegiate American football players.

*Research Implications*

Our results suggest GCs are ineffective at attenuating the linear and rotational head kinematics experienced by collegiate American football players; they may be effective when applied to other helmets. Hockey helmets for example are most commonly two thin layers of rigid padding covering the head, and therefore may benefit from an additional soft-shell covering such as the GC.

*Limitations*

This study had limitations related to the method of data collection chosen, as it may have been more beneficial to obtain data from games as opposed to practices, however, GCs are not allowed during games. Previous research has shown that the game concussion rate for collegiate American football players is 3.74 per 1000 athlete exposures, and 0.53 per 1000 athlete exposures (20). Comparing GC and non-GC data from American football games would allow us to see the effectiveness of GCs in a game setting. An additional limitation of this study is that behavioral differences while the players were wearing GCs was not considered. It is possible the players felt more or less protected with the additional padding and that changed the way they played. Our data did not find a significant change in total impacts from PRE to POST, however we cannot firmly conclude that there were no behavioral differences in the players practice style while wearing the GCs. Additionally, not all our players wore the same helmet underneath the GC, nor did we conduct an analysis to determine if a certain helmet pairs better with the GC to reduce the head kinematics experienced by the American football players. In addition, GCs are affixed mainly by Velcro straps and tend to slide off the players helmets during high contact drills or scrimmages. Though the workouts included in this study were video-verified and data would have been excluded had the GC come off completely, we cannot guarantee that the GC was not dislodged during some of the impacts included in this study. Lastly, no concussion rates were tracked during this study as the performance period was too short and that data may be more suited to a larger multi-team study to achieve sufficient power.

*Conclusion*

Consistent with the limited literature of Guardian Caps, our study found no reduction in PAA, PLA or total impacts when the Guardian Caps were affixed to traditional helmets in collegiate American football players. As significant head injuries are prominent in American football and other contact sports, with only a small literature base examining the effects of RHI, there is a need for continued research into improvements to contact sports equipment.

***Conflict of Interest***

None of the authors included on this project have any competing interests.

*Acknowledgements*

Funding was provided by the Neuroscience COBRE (P20GM103650), the Robert Z. Hawkins Foundation and the Nevada DRIVE Scholarship. We would like to thank all our participants who helped make this project possible. The results of this study are presented clearly and honestly, without any fabrication, falsification, or data manipulation. The results of this study do not constitute endorsement by ACSM.

*Figures/Tables*

**Table 1:** Frequency and percent of the entire sample of participants of each position and lineup designation, as well as average age at each practice timepoint

|  | Practice Number | PRE 1 | PRE 2 | PRE 3 | POST 1 | POST 2 | POST 3 |
|---|---|---|---|---|---|---|---|
| **Participants (n)** | | 13 | 12 | 17 | 14 | 14 | 13 |
| **Average age** | | 20 | 20 | 20 | 20 | 20 | 20 |
| **Position (n)** | | | | | | | |
| | OL | 3 (23.08%) | 2 (15.38%) | 2 (11.76%) | 2 (14.28%) | 3 (21.43%) | 3 (23.08%) |
| | DL | 2 (15.38%) | 1 (8.33%) | 2 (11.76%) | 1 (7.14%) | 2 (14.28%) | 2 (15.38%) |
| | RB | 2 (15.38%) | 1 (8.33%) | 3 (17.65%) | 3 (21.43%) | 3 (21.43%) | 2 (15.38%) |
| | TE | 3 (23.08%) | 2 (15.38%) | 3 (17.65%) | 2 (14.28%) | 1 (7.14%) | 3 (23.08%) |
| | LB | 3 (23.08%) | 6 (50%) | 7 (41.78%) | 6 (42.86%) | 5 (35.71%) | 3 (23.08%) |
| **Starters (n)** | | | | | | | |
| | Starter (S) | 2 (15.38%) | 0 (0%) | 1 (5.88%) | 0 (0%) | 1 (7.14%) | 2 (15.38%) |
| | Rotational (R) | 3 (23.08%) | 3 (25.0%) | 3 (17.65%) | 3 (21.43%) | 3 (21.43%) | 3 (23.08%) |
| | Non-Starter (N) | 7 (53.85%) | 6 (50.0%) | 9 (52.94%) | 8 (57.14%) | 6 (42.86%) | 7 (53.85%) |

Note: PLA = peak linear acceleration, PAA = peak angular acceleration, PRE = 3 practice sessions without the guardian cap and POST = 3 practice sessions with the guardian cap, OL = offensive lineman, DL = defensive lineman, RB = running back, TE = tight end, LB = linebacker

**Table 2:** Frequency and percent of the 7 consistent participants of each position and lineup designation, as well as average age at each practice timepoint

|  | Practice Number | PRE 1 | PRE 2 | PRE 3 | POST 1 | POST 2 | POST 3 |
|---|---|---|---|---|---|---|---|
| **Participants (n)** | | 7 | 7 | 7 | 7 | 7 | 7 |
| **Average age** | | 20 | 20 | 20 | 20 | 20 | 20 |
| **Position (n(%))** | | | | | | | |
| | OL | 1 (7.69%) | 1 (8.3%) | 1 (5.9%) | 1 (7.1%) | 1 (7.1%) | 1 (7.69%) |
| | DL | 1 (7.69%) | 1 (8.3%) | 1 (5.9%) | 1 (7.1%) | 1 (7.1%) | 1 (7.69%) |
| | RB | 1 (7.69%) | 1 (8.3%) | 1 (5.9%) | 1 (7.1%) | 1 (7.1%) | 1 (7.69%) |
| | TE | 1 (7.69%) | 1 (8.3%) | 1 (5.9%) | 1 (7.1%) | 1 (7.1%) | 1 (7.69%) |
| | LB | 3 (23.1%) | 3 (25%) | 3 (17.7%) | 3 (21.4%) | 3 (21.4%) | 3 (23.1%) |
| **Starter (n(%))** | | | | | | | |
| | Starter (S) | 0 (0%) | 0 (0%) | 0 (0%) | 0 (0%) | 0 (0%) | 0 (0%) |
| | Rotational (R) | 3 (23.1%) | 3 (25.0%) | 3 (17.7%) | 3 (21.4%) | 3 (21.4%) | 3 (23.1%) |
| | Non-starter (N) | 4 (30.8%) | 4 (33.3%) | 4 (23.5%) | 4 (28.6%) | 4 (28.6%) | 4 (30.8%) |

Note: PLA = peak linear acceleration, PAA = peak angular acceleration, PRE = 3 practice sessions without the guardian cap and POST = 3 practice sessions with the guardian cap, OL = offensive lineman, DL = defensive lineman, RB = running back, TE = tight end, LB = linebacker

**Table 3:** Mean(SD) values for PLA, PAA and Total Impacts for the entire sample and the 7 consistent players, as well as the significance (*p*-value)

| | PRE 1 | PRE 2 | PRE 3 | POST 4 | POST 5 | POST 6 | Sig. |
|---|---|---|---|---|---|---|---|
| **Full Sample** | | | | | | | |
| PLA (G) | 16.3(2.1) | 16.5(1.9) | 16.4(2.7) | 16.2(2.6) | 18.2(4.3) | 18.3(3.0) | 0.17 |
| PAA (rad/sec$^2$) | 970.8(191.0) | 1020.4(236.9) | 956.86(264.9) | 1064.4(297.2) | 1118.0(295.3) | 1017.1(244.5) | 0.50 |
| Total Impacts | 8.8(4.1) | 9.8(5.5) | 9.2(5.4) | 10.3(6.1) | 9.4(6.0) | 9.5(5.3) | 0.98 |
| **7 players** | | | | | | | |
| PLA (G) | 15.7(1.2) | 16.8(2.1) | 16.2(2.4) | 15.8(2.7) | 17.2(4.5) | 18.6(3.1) | 0.67 |
| PAA (rad/sec$^2$) | 927.6(195.5) | 1000.3(203.6) | 925.6(155.2) | 955.8(139.4) | 955.5(159.8) | 1202.2(370.0) | 0.80 |
| Total Impacts | 9.7(3.0) | 9.7(6.7) | 9.4(5.2) | 10.9(6.5) | 7.7(5.6) | 9.3(6.2) | 0.82 |

Note: PLA = peak linear acceleration, PAA = peak angular acceleration, PRE = 3 practice sessions without the guardian cap and POST = 3 practice sessions with the guardian cap.

**Table 4:** Collapsed mean(SD) values for PLA, PAA and Total Impacts for the entire sample and the 7 consistent players, as well as the significance (*p*-value)

|  | **Variable** | **PRE** | **POST** | ***p*-value** | **Cohen's d** |
|---|---|---|---|---|---|
| **Full Sample** | | | | | |
| | PLA (G) | 16.3(1.99) | 17.2(3.27) | 0.20 | 0.33 |
| | PAA (rad/sec$^2$) | 992.1(209.17) | 1029.4(261.01) | 0.51 | 0.16 |
| | Total Impacts | 9.2(4.69) | 9.7(5.65) | 0.72 | 0.09 |
| **7 players** | | | | | |
| | PLA (G) | 16.2(1.18) | 17.2(2.79) | 0.32 | 0.47 |
| | PAA (rad/sec$^2$) | 951.2(95.40) | 1038.0(166.81) | 0.29 | 0.64 |
| | Total Impacts | 9.6(4.15) | 9.7(5.04) | 0.88 | 0.02 |

Note: PLA = peak linear acceleration, PAA = peak angular acceleration, PRE = 3 practice sessions without the guardian cap and POST = 3 practice sessions with the guardian cap.

**Table 5:** Significance (p-values) for pairwise comparisons of PAA, PLA and total impacts for the 7 consistent players

| | PAA | | | |
|---|---|---|---|---|
| | PRE 1 | PRE 2 | POST 1 | POST 2 |
| PRE 2 | 0.559 | | | |
| PRE 3 | 0.985 | 0.461 | | |
| POST 2 | | | 0.927 | |
| POST 3 | | | 0.148 | 0.096 |
| | **PLA** | | | |
| | PRE 1 | PRE 2 | POST 1 | POST 2 |
| PRE 2 | 0.387 | | | |
| PRE 3 | 0.638 | 0.582 | | |
| POST 2 | | | 0.331 | |
| POST 3 | | | 0.074 | 0.395 |
| | **Total Impacts** | | | |
| | PRE 1 | PRE 2 | POST 1 | POST 2 |
| PRE 2 | 1.000 | | | |
| PRE 3 | 0.904 | 0.876 | | |
| POST 2 | | | 0.343 | |
| POST 3 | | | 0.616 | 0.313 |

Note: PLA = peak linear acceleration, PAA = peak angular acceleration, PRE = 3 practice sessions without the guardian cap and POST = 3 practice sessions with the guardian cap.

**Table 6:** Significance (p-values) for pairwise comparisons of PAA, PLA and total impacts for across the entire sample

| | PAA | | | |
|---|---|---|---|---|
| | PRE 1 | PRE 2 | POST 1 | POST 2 |
| PRE 2 | 0.562 | | | |
| PRE 3 | 0.294 | 0.690 | | |
| POST 2 | | | 0.382 | |
| POST 3 | | | 0.224 | 0.835 |
| | **PLA** | | | |
| | PRE 1 | PRE 2 | POST 1 | POST 2 |
| PRE 2 | 0.877 | | | |
| PRE 3 | 0.167 | 0.198 | | |
| POST 2 | | | 0.100 | |
| POST 3 | | | 0.938 | 0.158 |
| | **Total Impacts** | | | |
| | PRE 1 | PRE 2 | POST 1 | POST 2 |
| PRE 2 | 0.490 | | | |
| PRE 3 | 0.243 | 0.795 | | |
| POST 2 | | | 0.698 | |
| POST 3 | | | 0.830 | 0.849 |

Note: PLA = peak linear acceleration, PAA = peak angular acceleration, PRE = 3 practice sessions without the guardian cap and POST = 3 practice sessions with the guardian cap.